# On the similarity between Nakagami-m Fading distribution and the Gaussian ensembles of random matrix theory

Sherif M. Abuelenin

Department of Electrical Engineering, Faculty of Engineering, Port-Said University

*Abstract*

We report the similarity between the Nakagami-m fading distribution and the three Gaussian ensembles of random matrix theory. We provide a brief review of random matrix theory and wireless fading. We show that the Nakagami-m distribution serves as mapping between the three ensembles. The statistics of the wireless fading amplitude, as modeled by Nakagami-m distribution, provide a rare example of a classically chaotic system that exhibits a transition between the Gaussian orthogonal, Gaussian unitary, and Gaussian symplectic ensembles of random matrix theory.

*Index Terms*: Gaussian ensembles, Nakagami-m distribution, Random matrix theory, Rapid fading, Wireless channel.

I.  INTRODUCTION

In wireless communications, small-scale fading is the short-term variation in the received signal amplitude due to the constructive and destructive interference of the multiple signal paths [1]. When studying small-scale fading in wireless communications, several statistical models exist to describe the probability distribution of the amplitude of the received signal. The simplest is Rayleigh fading distribution which assumes that there are many statistically independent multipath components with random amplitudes and delays. The instantaneous amplitude is obtained as the modulus of the random process; $r(t) = X(t) + jY(t)$, where *X* and *Y* are uncorrelated zero mean Gaussian processes with equal variances. This will be discussed in more details below. Unlike most fading distributions that model certain conditions, the Nakagami-m [2]-[4] fading model is capable of modeling a wide class of fading channel conditions and it fits well the empirical data [5].

Electromagnetic wave propagation in complex environments can be considered a chaotic system [6]. When the wavelength is short compared to the typical length scale of the system, the wave behavior can be approximated by ray equations. Ray tracing is currently accepted as an accurate method of studying electromagnetic propagation in high frequency regimes [8]. Microwave signals used in communication systems commonly have wavelengths between few centimeters to one meter, which are smaller than objects encountered in a typical outdoor propagation environment (e.g. buildings, vehicles, trees, etc.). Ray trajectories in such complex environments show chaotic behavior [6]. Therefore, it was asserted [6], [7] that such systems are expected to show universal statistical properties, and therefore cam be modeled using random matrix theory (RMT) tools. RMT provides a framework for describing the statistical properties of spectra for quantum systems whose classical counterpart is chaotic [9], [10]. RMT was also successfully applied in a variety of problems that include classical chaotic systems [11], [12].

Here, we report the resemblance between the Nakagami-m fading model, and the three Gaussian ensembles of the RMT. Related work is reviewed in the next section. Small-scale fading and Nakagami-m fading model are discussed the section III. A discussion on RMT and the Gaussian ensembles is provided in section IV. The paper is concluded with a summary and discussion of the reported findings.

## II. RELATED WORK

The earliest applications of RMT in wireless communications date back to the pioneering works of Foschini [13] and Telatar [14] on characterizing the capacity of multi-antenna fading channels. A 2004 survey [15] provided a survey on RMT in wireless communications. There are several other examples of more recent applications in wireless communications spectrum sensing [16]-[19]. But it was only recently that RMT was directly applied, by J-Yeh et. al [6], [7], in studying the wireless fading amplitude. They derived an RMT-based model of wireless fading in certain regimes and included both Rayleigh [20] and Rician [21] fading. This RMT modeling was based on the chaotic nature of electromagnetic wave propagation in complex environments [6]. When the wavelength is short compared to the typical length scale of the system, the wave behavior can be approximated by ray equations. In wireless communication, the ray limit is attained when the wavelength is small in comparison to size of the scattering system. Such chaotic systems are expected to show universal statistical properties (see [6] and references therein). The model was verified using experimental setup utilizing chaotic microwave cavities that was based on earlier works [22]-[25] where GOE-GUE transition was observed by breaking the time-reversal invariance inside the cavity. In the earlier results, the agreement between experimental and RMT generated results supported the use of RMT to model statistics of real semi-classical wave-chaotic systems [25]. Within the RMT framework, microwave cavities have been used to simulate quantum billiards [12]. This utilization was based on the similarity between the electric field wave equation inside the cavity and Schrödinger's equation for a two-dimensional billiard. Inside microwave cavities, the ray limit is attained if the electromagnetic wavelength is small compared to the dimension of the cavity [6], [12].

Additionally, it was known for a while that the mathematical form of the Wigner surmise for the GOE distribution is identical with Rayleigh's distribution [26]. Kumar and Pandey [27] observed the similarity between the Nakagami-q (a.k.a. Hoyt) fading distribution and the Laguerre ensembles of RMT. Nakagami-q distribution is an approximation of Nakagami-m distribution that spans from single-sided Gaussian to Rayleigh distributions, as $q$ changes from 0 to 1. It is used to describe fading in multi-input multi-output (MIMO) wireless communications. Kumar and Pandey reported that the Nakagami-q distribution corresponds to a random matrix ensemble which interpolates between the Laguerre orthogonal ensemble (LOE) and the Laguerre unitary ensemble (LUE). The LOE case occurs when the variance of either the real or complex component is zero, whereas the LUE (Rayleigh) case is obtained, as stated earlier, for equal variances. These authors commented on the findings that it would be of interest to extend their results to deal with the case of correlated channels.

### III. Nakagami-m Fading

The empirical Nakagami-m distribution was first proposed in the 1940's by Minoru Nakagami [4] to model small-scale (rapid) fading in long-distance wireless channels. Small-scale fading is the short-term variation in the received signal amplitude due to the constructive and destructive interference of the multiple signal paths between the transmitter and receiver [1].

The propagation of electromagnetic waves between a transmitter and a receiver through a complex environment is accompanied by typical wave phenomena such as diffraction, scattering, reflection, and absorption [28]. Therefore, the received signal is composed of various components with different delays and amplitudes; these commonly include a direct component and scattered and reflected components. The total received electric field can be interpreted as a vector sum in the complex plane [29].

Due to the existence of a great variety of fading environments, several statistical models are used to describe the probability distribution of the received signal amplitude. Small-scale fading models include Rayleigh [20], Rice [21], Nakagami-m [4], Hoyt [4], [30] and Weibull [31] distributions [32].

Rayleigh distribution is the simplest fast-fading model. It is based on the assumption that there are a large number of statistically independent reflected and scattered components with random amplitudes and delays (i.e. pure diffuse scattering). The instantaneous amplitude is obtained as the modulus of a complex Gaussian process; $r(t) = X(t) + jY(t)$, where X and Y are uncorrelated zero mean Gaussian processes with equal variances.

In general, other fading distributions can be theoretically derived by assuming that the random vector phases are not uniformly distributed, or that the vectors are correlated [33] (for summary, see [34]). For example, when there exists a certain communication path

(usually, a line-of-sight) that results in a strong component besides the multiple independent multipath components, Rician distribution provides a better model [1].

Making no assumptions on the statistics of the amplitudes and phases of the multiple received versions (i.e., allowing X and Y to have different variances or being correlated [35]) led to the more general Nakagami-m distribution. Nakagami (and independently, Beckmann) has derived the m-distribution as an approximate form of the distribution of the sum of large number of vectors allowing correlated components with different mean values and variances [36]. The Nakagami-m distribution has gained a lot of attention due to its ability to model a wide class of fading channel conditions and to fit well the empirical data [5]. It was originally proposed because it matched empirical data better than other distributions (i.e. Rayleigh, Rice, or lognormal distributions [16], [37]). This was confirmed by measurements as reported in [38], [39]. The probability density function of the distribution is given by (1).

$$f(x) = \frac{2m^m}{\Gamma(m)\Omega^m} x^{2m-1} \exp\left(-\frac{m}{\Omega}x^2\right) u(x); \ m \geq 0.5, \ \Omega \geq 0. \qquad (1)$$

where $\Gamma(m)$ is the Gamma function, and $u(x)$ is the Heaviside unit step function. The distribution has two positive-valued parameters, $m$ and $\Omega$. The fading parameter $m$ controls the shape of the distribution, and $\Omega$ controls the spread. The value of the parameter $m$ serves as an indicator of the fading severity. The minimum value that $m$ can obtain equals 0.5 [4], in this case the distribution is equivalent to a single-sided Gaussian distribution and represents the most severe fading that can be modeled by the Nakagami-m distribution [56]. When $m = 1$, the Nakagami-m distribution becomes equivalent to Rayleigh distribution. Larger values of $m$ indicate the existence of a strong (normally, a line-of-sight) component, which represent less-severe fading conditions. Least severe fading is obtained as $m$ approaches infinity [40]. Also, Nakagami-m model serves as an approximation to the Rician fading distribution and is used for analytical simplicity and ease of manipulation [41].

As mentioned above, Nakagami also introduced the Nakagami-q distribution (a.k.a. Hoyt distribution). Nakagami-q distribution is used to model the fading channels when it is more severe than the Rayleigh fading. The value of the fading parameter q ranges from 0 to 1. For q = 0, the distribution is equivalent to a single-sided Gaussian fading, the highest severity that can be modeled by Hoyt distribution. And for q = 1, the distribution is equivalent to Rayleigh fading, representing the lowest severity. It should be noted that Nakagami-q distribution is a special case of the more general Nakagami-m distribution.

## IV. RANDOM MATRIX THEORY

Random matrix theory (RMT) was initially proposed for describing the statistical properties of spectra of complex many-body quantum systems [11], [12]. RMT models the Hamiltonian of the system by an ensemble of random matrices that depends only on the symmetry properties of the system. Time-reversal-invariant chaotic system is represented by a Gaussian orthogonal ensemble (GOE) when the system has rotational symmetry and by a Gaussian symplectic ensemble (GSE) otherwise [10]. Systems without time-reversal invariance are represented by the Gaussian unitary ensemble (GUE).

RMT was initially proposed by Wigner [42] in studying the statistical properties the neutron excitation spectra of heavy nuclei [43]. According to Wigner, these are many-particle systems whose interaction is so complex that their Hamiltonian should behave like a random matrix [43]. Wigner conjectured (surmised) [44] that the statistical distribution of the distances between neighboring energy level should be given by the following expression for the GOE [11];

$$p_\beta(s) = \frac{\pi s}{2} \exp\left(-\frac{\pi}{4} s^2\right), \qquad s = \frac{S}{D} \qquad (2)$$

where the variable $s$ is the actual spacing $S$ normalized by the mean level spacing $D$ [12]. This parameter-free expression was analytically obtained later [45], and it corresponds to the nearest-neighbor spacing distributions (NNSD) of ordered eigenvalues of ensembles of 2×2 matrices. It also presents close approximation to the exact results for the case of large-sized random matrices. Similar expressions were obtained for the cases of GUE and GSE [45], [46].

It has been well established that random matrix statistics are considered as experimental signatures of quantum chaos, however, this relation is not fully theoretically understood [47]. Later, RMT was successfully applied in a variety of problems that are quite different from complex many-body systems [12]. Examples include number theory, networks, traffic, biology, and classical chaotic systems [11], [12]. The continuing discovery of various applications of RMT is said to give a strong indication that eigenvalues of random matrices provide a "fundamental model for sequences of dependent random numbers" [43]. A complete understanding of this universality [48] is still not reached [43].

### A. Gaussian ensembles

As stated above, RMT describes quantum systems whose classical counterparts are chaotic [49] and correctly predicts the strong short-range correlations of the eigenvalues due to interactions. The generalized Wigner surmise [42] of RMT that describes the NNSD of eigenvalues is given by

$$p_\beta(s) = a_\beta s^\beta \exp(-b_\beta s^2) \tag{3}$$

where $s$ is the normalized spacing, $\beta$ is known as the 'level repulsion' parameter [55], and the quantities $a_\beta$ and $b_\beta$ are chosen such that;

$$\int_0^\infty p_\beta(s)ds = 1 \text{ and } \int_0^\infty p_\beta(s)sds = 1$$

This leads to;

$$a_\beta = 2\frac{\left[\Gamma\left(\frac{\beta+2}{2}\right)\right]^{\beta+1}}{\left[\Gamma\left(\frac{\beta+1}{2}\right)\right]^{\beta+2}} \text{ and } b_\beta = \frac{\left[\Gamma\left(\frac{\beta+2}{2}\right)\right]^2}{\left[\Gamma\left(\frac{\beta+1}{2}\right)\right]^2}$$

This density depends only on the choice of the matrix ensemble [47], with real, complex, and quaternionic matrices leading to $\beta$ = 1, 2, 4, corresponding to the GOE, GUE, and GSE ensembles of RMT, respectively. This yields, as derived for the case of 2x2 matrices [46];

$$p(s) = \begin{cases} (s\pi/2)\exp(-s^2\pi/4), & \text{GOE} \\ (32s^2/\pi^2)\exp(-s^2 4/\pi), & \text{GUE} \\ (2^{18}s^4/3^6\pi^3)\exp(-s^2 64/9\pi), & \text{GSE} \end{cases}$$

Unlike the GOE statistics, examples of the two other universality classes are much rarer [50]. GUE statistics was observable in certain systems, including microwave cavities with ferrite strips [6], [7]. There are only few examples where transitions from GOE to GUE statistics in dependence of a parameter of the system could be studied (see [50] and references therein). GSE examples are even rarer, and microwave cavity realization of GSE systems was only reported lately [51].

  *B. Nakagami-m distribution and the Gaussian ensembles:*

The three Gaussian ensembles represented in the previous equations are exactly the Nakagami-m distribution for m = 1, 1.5, 2.5 (the corresponding values for $\Omega$ are $\pi/4$, $3\pi/8$, $25\pi/18$, respectively). In other words, $\beta = 2m - 1$. And by setting the mean value of the Nakagami-m distribution to 1;

$$\Omega = \frac{m[\Gamma(m)]^2}{[\Gamma(m+0.5)]^2}$$

In real long distance wireless communication scenarios, estimating the value of the parameter *m* resulted in values ranging from around one to more than five [52]-[54]. E.g. in highway communications, where vehicles equipped with transmitters and receivers, Ref [53] reports that the value of *m* varied between one and three, depending on the distance *d* between the transmitter and the receiver (1 for *d* > 150, 1.5 for 150 > *d* > 50, and 3 for 50 > *d*). This implies that the statistics of the wireless fading in such environment represent a transition consecutively between the GOE, GUE, and GSE as the distance between the transmitter and receiver decreases.

Therefore, the Nakagami-m distribution represents an interpolation between the three Gaussian ensembles of RMT. This is a rare example of a classical chaotic system that encounters a transition between the three Gaussian ensembles of RMT.

V. SUMMARY

The discussion presented above shows that the empirical Nakagami-m distribution, which models various wireless fast-fading conditions, corresponds to a mapping between the three Gaussian ensembles of RMT. It has the same functional form of the (general) Wigner surmise. The parameter $\beta$, which indicates the level of repulsion in the Gaussian ensembles, is directly related to the m parameter of the Nakagami-m distribution, which defines the relative strength of the LOS component in the wireless channel.

The long-distance wireless fading channel, which can be considered a classical chaotic system, provides a rare example of a system that encounters GOE-GUE-GSE transitions.

Unlike the GOE statistics, examples of physical systems that exhibit the two other universality classes are much rarer [50]. There are only few examples where transitions from GOE to GUE statistics in dependence of a system parameter could be studied [50]. The long distance wireless communication channel fading amplitude, when described by the Nakagami-m distribution, shows a very rare example of a physical system that exhibits a gradual transition between the three classes. In this case, the system parameter controlling the transition is the severity of the fading amplitude, as identified by the parameter *m*.